\documentclass[aps,prb,reprint,amssymb,amsmath,superscriptaddress]{revtex4-1}

\usepackage{mathtext}
\usepackage{amssymb}
\usepackage{graphicx}
\usepackage{amsmath}
\begin{document}
\title{Dephasing and interwell transitions in double quantum well heterostructures}

\author{G.~M.~Minkov}
\affiliation{Institute of Metal Physics RAS, 620990 Ekaterinburg,
Russia}

\affiliation{Ural State University, 620083 Ekaterinburg, Russia}

\author{A.~V.~Germanenko}

\author{O.~E.~Rut}
\affiliation{Ural State University, 620083 Ekaterinburg, Russia}

\author{A.~A.~Sherstobitov}
\affiliation{Institute of Metal Physics RAS, 620990 Ekaterinburg,
Russia}

\affiliation{Ural State University, 620083 Ekaterinburg, Russia}

\author{A.~K.~Bakarov}
\author{D.~V.~Dmitriev}

\affiliation{Institute of Semiconductor Physics RAS, 630090
Novosibirsk, Russia}

\date{\today}

\begin{abstract}
The interference quantum correction  to the conductivity in the gated
double quantum well Al$_x$Ga$_{1-x}$As/GaAs/Al$_x$Ga$_{1-x}$As
structures is studied experimentally. The consistent analysis of the
interference induced positive magnetoconductivity  allows us to find
the interwell transition time $\tau_{12}$ and the electron dephasing
time $\tau_\phi$. It has been obtained that $\tau_{12}^{-1}$ resonantly
depends on the difference between the electron densities in the wells
as predicted theoretically. The dephasing times have been determined
under the conditions when one and both quantum wells are occupied. The
surprising result is that the $\tau_\phi$ value in the one well does
not depend on the occupation of the other one.
\end{abstract} 

\maketitle

\section{Introduction}
\label{sec:intr}

The quantum corrections to the conductivity of the two-dimensional (2D)
electron gas with the single-valley parabolic energy spectrum is
extensively investigated last three decades (see
Refs.~\onlinecite{AA85,FinRev,Alei99,Zala01,Gor04,Punnoose05,Minkov09,Min04-4,Minkov07-1}
and references therein). A reasonably good agreement between
experimental and theoretical results evident at relatively high
conductivity, $\sigma\gtrsim (10-15)\,G_0$, where $G_0=e^2/\pi\, h$,
attests adequate understanding of the role of the quantum interference
and electron-electron ({\it e-e}) interaction in the transport
properties of 2D systems. The quantum corrections in the double quantum
well structures are studied significantly
less.\cite{Averkiev98,Raichev00,Min00-3,Pagnossin08} Specific features
of the corrections in this case are governed by relationship between
the following parameters: the transport times $\tau$, the phase
relaxation time $\tau_\phi$, and the interwell transition time
$\tau_{12}$, the temperature length $L_T=\sqrt{D/T}$ (where $D$ is the
diffusion coefficient), the interwell distance $d$, and the screening
length $r_0$, which is half the Bohr radius, $r_0=a_B/2$. In the
limiting cases such structures can look like a single quantum well
structure (at $\tau_{12}\ll \tau_\phi,\, 1/T$) or a structure with two
uncoupled wells (at $\tau_{12}\gg \tau_\phi,\, 1/T$ and $d\gg r_0$). In
the last case, the quantum corrections to the conductivity are simply
the sum of the corrections of each well, however, the crossover to this
case for the WL and interaction corrections may occur at different
conditions.

The intermediate cases are more interesting and diversified. Concerning
the interaction correction, the gain of the interaction in a multiplet
channel can be so significant that it can lead to the change of the
sign of the temperature dependence of the conductivity from dielectric
($d\sigma/dT>0$) to metalliclike ($d\sigma/dT<0$) as it takes place in
the 2D systems with double valley energy spectrum. The interference
correction, even without interwell transitions ($\tau_{12}\gg
\tau_\phi$), can differ from the case of non-interacting wells due to
inelasticity of the {\it e-e} interaction of carriers in the different
wells. This interaction can change the dephasing rate significantly.
Just this effect in the gated double quantum well
Al$_{x}$Ga$_{1-x}$As/GaAs/Al$_x$Ga$_{1-x}$As structures is studied in
the present paper. Analyzing the interference induced positive
magnetoconductivity we show that the interwell transition rate
resonantly depends on the gate voltage. We also show that the dephasing
rate is almost insensitive to the number of the wells occupied contrary
to the theoretical expectations.

\section{Experimental details}
\label{sec:exp} 
\begin{figure}
\includegraphics[width=0.8\linewidth,clip=true]{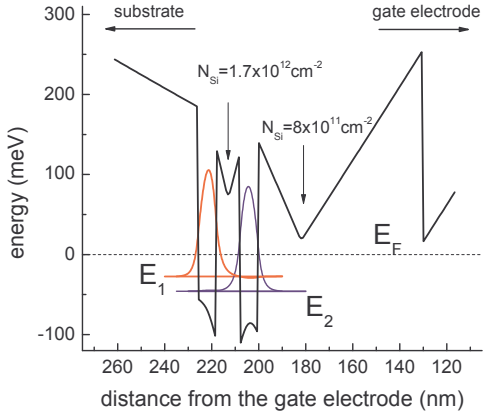}
\caption{(Color online) The energy diagram of the structure 3243 for $V_g=0$.
}\label{f1}
\end{figure}

Two structures  grown by the molecular beam epitaxy on
a semiinsulating GaAs substrate were investigated. The first structure
3243 consists of $200$~nm-thick undoped GaAs buffer layer, $50$~nm
undoped Al$_{0.3}$Ga$_{0.7}$As layer,  two $8$~nm GaAs quantum wells
separated by $10$~nm Al$_{0.3}$Ga$_{0.7}$As barrier, $70$~nm
Al$_{0.3}$Ga$_{0.7}$As undoped layer. The thickness of the GaAs cap
layer is $130$~nm. The main doping $\delta$ layer of Si is situated in
the barrier center. Because the gate voltage in such a type of
structures can effectively decrease the electron density only, an
additional Si $\delta$ layer is located above the upper quantum well at
distance of $18$~nm from the well interface. The energy diagram of the
structure 3243 is shown in Fig.~\ref{f1}. The second structure 3154
differs by the doping level and has lower electron density and
mobility. Such structure design allows us to have the close mobilities
in the wells at equal electron density that is very important for
reliable determination of the phase relaxation time as will be
discussed below.  The samples were mesa etched into standard Hall bars
and then an Al gate electrode was deposited by thermal evaporation onto
the cap layer through a mask. All the measurements were carried out in
the Ohmic regime using DC technique. The results obtained were mostly
analogous and we will discuss more thoroughly the results obtained for
the structure 3243.

\section{Results and discussion}
\label{sec:results}

The shape of magnetoconductivity $\Delta\sigma(B)$  caused by
suppression of the weak localization in the transverse magnetic field
$B\parallel n$, where $n$ is the normal to the structure plane, is
widely varied in the double well structures. Really, even in the case
of low interwell transition rate, $\tau_{12}^{-1}$, when
$\Delta\sigma(B)$ is barely the sum of the contributions from each well
\begin{eqnarray}
 \frac{\Delta\sigma(B)}{G_0}&=&  \frac{\rho_{xx}^{-1}(B)-\rho_{xx}^{-1}(0)}{G_0} \nonumber \\
 &=& \alpha_1{\cal H}\left(\frac{\tau_{\phi 1}}{\tau_1}\frac{B}{B_{tr1}}\right)
 +\alpha_2{\cal H}\left(\frac{\tau_{\phi 2}}{\tau_2}\frac{B}{B_{tr2}}\right)  ,
\label{eq10}
\end{eqnarray}
it depends, as seen, on the large number of parameters: the two phase
relaxation times, $\tau_{\phi 1}$ and $\tau_{\phi 2}$, the two
transport times, $\tau_1$ and $\tau_2$, and two transport magnetic
fields, $B_{tr1}=\hbar/4eD_1\tau_1$ and $B_{tr2}=\hbar/4eD_2\tau_2$.
The remaining designations in Eq.~(\ref{eq10}) are the following:
\begin{equation}
{\cal
H}(x)=\psi\left(\frac{1}{2}+\frac{1}{x}\right)+\ln{x},
 \label{eq20}
\end{equation}
where $\psi(x)$ is a digamma function, $\alpha_{1}$ and $\alpha_{2}$
are prefactors, which appear due to not rigorous fulfilment of the
diffusion approximation $\tau_i/\tau_{\phi i}\ll 1$,\cite{Min00-1} due
to the magnetic field dependence of the dephasing time\cite{Germ07} and
the contributions of the second order corrections.\cite{Min04-2}
Obviously it is impossible to find six parameters fitting the single
smooth curve. Reliable obtaining of the phase relaxation time is
possible only when the parameters of the wells $\tau_i$,  $\tau_{\phi
i}$, and $B_{tr\,i}$ are close to each other. Therefore, before to
inspect interference induced low field magnetoconductivity let us
analyze the transport in high magnetic field in more detail.

\begin{figure}
\includegraphics[width=\linewidth,clip=true]{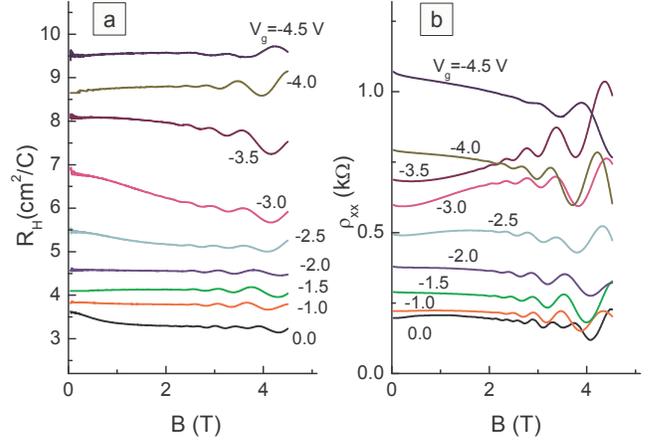}
\caption{(Color online) The magnetic field dependences of $R_H$ (a) and $\rho_{xx}$ (b) for
different gate voltages, $T=4.2$~K.
}\label{f2}
\end{figure}

\begin{figure}
\includegraphics[width=\linewidth,clip=true]{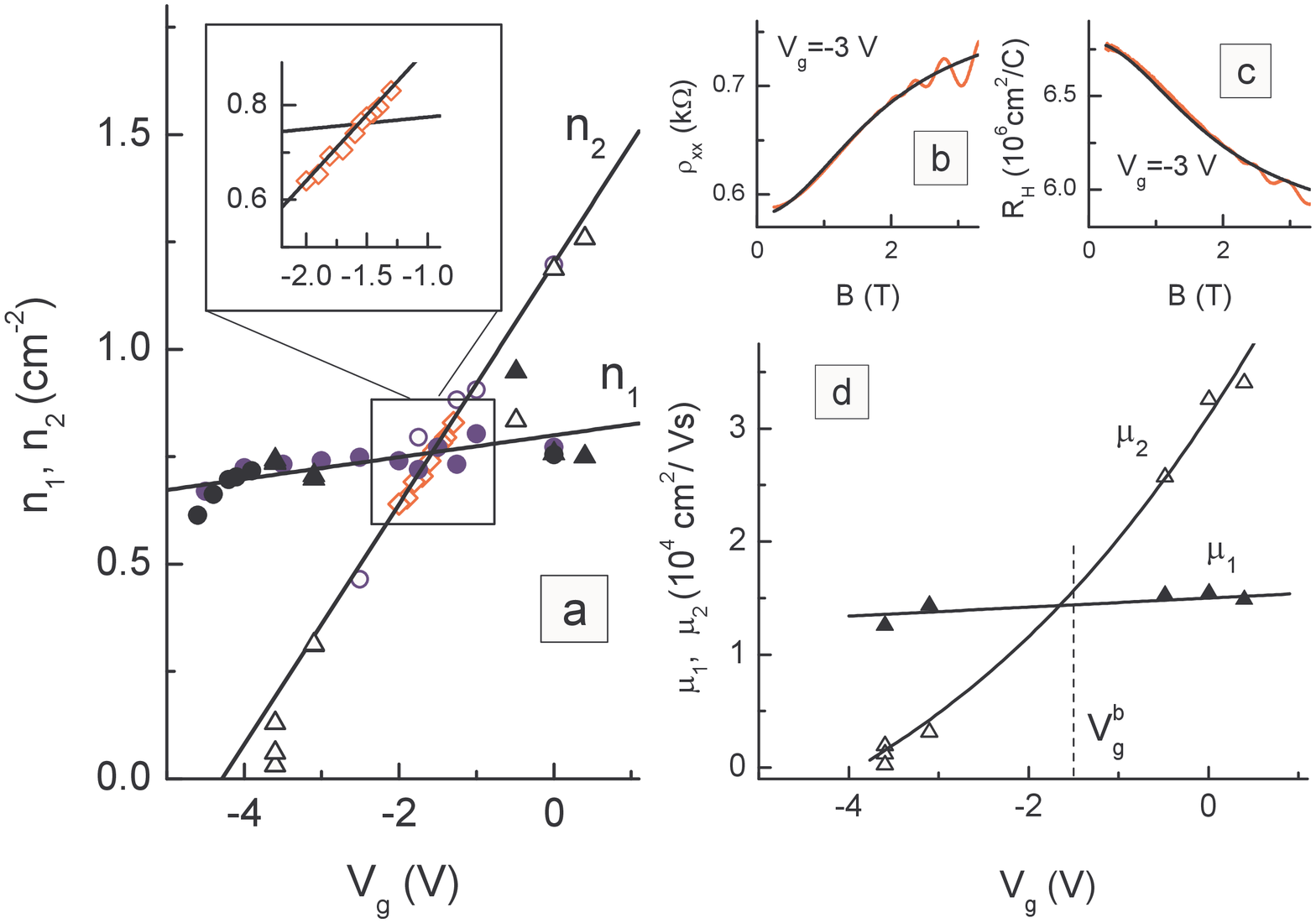}
\caption{(Color online) (a) The electron densities in the wells
plotted against the gate voltages. The circles and diamonds are obtained from the SdH
oscillations analysis, the triangles are obtained from the analysis of the monotonic
run of the $\rho_{xx}$~vs~$B$ and $R_H$~vs~$B$ curves.
(b) and (c) Examples of the fit of the experimental curves by the classical formula for
two types carriers transport for $V_g=-3.0$~V. (d) The gate voltage dependence of
the mobilities  in the wells obtained from the classical treatment of $\rho_{xx}$~vs~$B$ and $R_H$~vs~$B$
data.
}\label{f3}
\end{figure}

The magnetic field dependences of the Hall coefficient
$R_H=\rho_{xy}/B$ and transverse resistivity $\rho_{xx}$ for the
different gate voltages are presented in Fig.~\ref{f2}. First we
analyze the Shubnikov-de Haas (SdH) oscillations. As seen the
oscillations picture is rather complicated at some gate voltages that
stems from the difference in the electron densities in the wells.  The
$V_g$ dependences of the $n_1$ and $n_2$ values obtained from the
Fourier analysis are shown in Fig.~\ref{f3}(a) by circles. One can see
that $n_1$ and $n_2$ linearly decrease with the $V_g$ decrease.
Therewith, the decrease rate for one of the wells at $V_g\gtrsim -4$~V
is about fifteen times larger than that for other one: $dn_2/dV_g\simeq
2.8\times 10^{11}$~cm$^{-2}$/V against $dn_1/dV_g\simeq 2\times
10^{10}$~cm$^{-2}$/V. Both values are in agreement with the results of
simple estimations. The  rate $dn_2/dV_g$ agrees well with the
geometric capacity between the upper well and the gate electrode. The
nonzero value of the rate  for the lower well, $dn_1/dV_g$, results
from the finite compressibility  of the electron gas in the upper well.
Simple estimation gives the decrease rate due to this effect about
$2.3\times 10^{10}$ cm$^{-2}$/V that is close to the experimental
$dn_1/dV_g$ value in Fig.~\ref{f2}(a).

At $V_g\lesssim -4$~V, the upper well is fully depleted and no longer
screens the gate electric field.  Therefore $dn_2/dV_g$ is enhanced up
to approximately $2.5\times 10^{11}$ cm$^{-2}$/V, which agrees with the
geometric capacity again. Thus, the analysis of the SdH oscillations
shows that the structure comes to the balance, i.e., to the state when
the electron densities in the wells become close to each other, at
$V_g^b=-(1.5\pm 0.1)$~V.

As described above, to obtain $\tau_\phi$ reliably, it is very
important, besides the equal electron densities in the wells, to have
the close values of the transport magnetic fields, i.e., the close
values of the mobility. We used different ways to determine the
mobilities. First we analyzed the monotonic run of the $\rho_{xx}$ vs
$B$ and $R_H$ vs $B$ curves within the framework of the model of the
classical transport by two types of carriers. An example of such the
data processing is shown in Fig.~\ref{f3}(b) and Fig.~\ref{f3}(c). The
values of the electron densities in the wells obtained by this method
[presented in Fig.~\ref{f3}(a) by triangles] agree well with the that
found from the SdH oscillations. Unfortunately, this method gives good
results in the case when the mobilities differ noticeably. For the
structure 3243 it happens  at $V_g\simeq -(0\ldots0.5)$~V and
$V_g\simeq-(3\ldots3.5)$~V [see Fig.~\ref{f3}(d)]. By assuming that
$\mu_1$ and $\mu_2$ depend on the gate voltage monotonically, one
obtains that the mobilities differ not so strongly in the balance:
$\mu_1\simeq\mu_2= (1.5\pm 0.1)\times 10^4$~cm$^2$/Vs at
$V_g^b=-1.5$~V.

\begin{figure}
\includegraphics[width=\linewidth,clip=true]{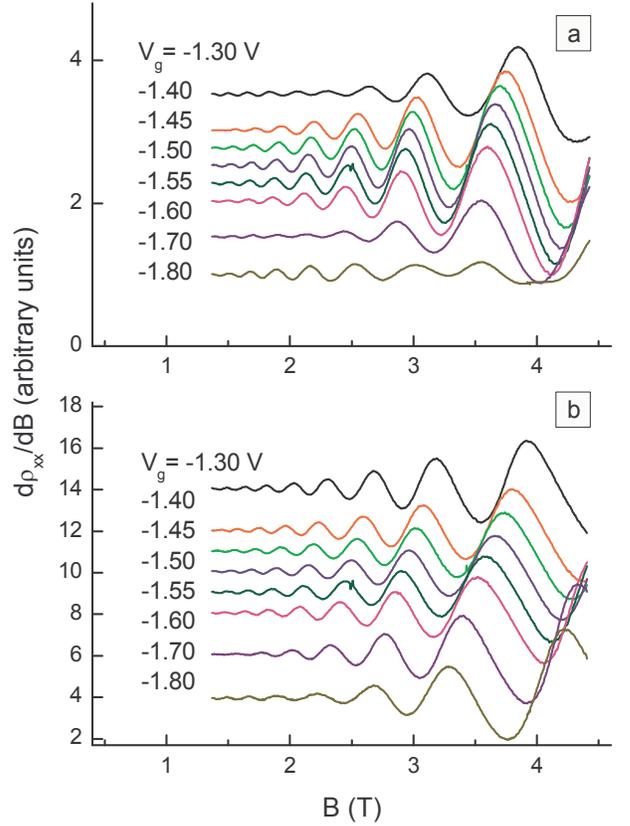}
\caption{(Color online) The SdH oscillations of $d\rho_{xx}/dB$ measured experimentally
(a) and extracted for the upper quantum well as described in the text (b).
}\label{f4}
\end{figure}

Another way to insure that the mobilities in the wells are close in the
magnitude at  the balance is the detailed analysis of the SdH
oscillations  [see Fig.~\ref{f4}(a)]. If one neglects the tunneling
splitting, $\Delta_\text{SAS}$, as compared with the broadening due to
scattering, $\hbar/\tau_q$, the conductivity tensor for the structure
can be represented as the sum of that for each well:
\begin{equation}
\sigma_{j}(B,V_g)=\sigma_{j}^{(1)}(B,V_g)+\sigma_{j}^{(2)}(B,V_g),\,\, j=xx,\,xy.
 \label{eq30}
\end{equation}
For our case the estimation gives $\Delta_{\text{SAS}}\simeq 0.1$ meV,
that is much less than $\hbar/\tau_q\simeq 5$~meV found from the $B$
dependence of the SdH oscillations amplitude. So, such the approach
seems reasonable. For the case of $\mu_1=\mu_2$ in the balance:
$\sigma_{j}^{(1)}(B,V_g^b)=\sigma_{j}^{(2)}(B,V_g^b)=\sigma_{j}(B,V_g^b)/2$.
Since the electron density and mobility in the lower well are
practically independent of $V_g$ near the balance [see
Fig.~\ref{f3}(d)], the components $\sigma_{j}^{(2)}$ for the upper well
can be extracted as
\begin{equation}
\sigma_{j}^{(2)}(B,V_g)=\sigma_{j}(B,V_g)-\frac{1}{2}\,\sigma_{j}(B,V_g^b).
 \label{eq40}
\end{equation}
The components in the right hand side of Eq.~(\ref{eq40}) were found
from the measured resistivity components $\rho_{xx}$ and $\rho_{xy}$.
After obtaining $\sigma_{xx}^{(2)}$ and $\sigma_{xy}^{(2)}$ we have
calculated $\rho_{xx}^{(2)}$ and plotted it in Fig.~\ref{f4}(b).
Therewith, to remove the monotonic component, the $\rho_{xx}$ vs B
curves were differentiated. One can see that: (i) the oscillations
extracted consist of one period; (ii) the $n_2$ vs $V_g$ dependence
found from the Fourier analysis is linear [see insert in
Fig.~\ref{f3}(a)]; (iii) its slope is close to that found within whole
$V_g$ range; (iv) the oscillations amplitude is practically independent
of $V_g$ while $V_g=-(1.7\ldots 1.3)$~V. Such the behavior can be
observed only when the mobilities in the wells at the balance,
$V_g=-1.5$~V, are close to each other.

\begin{figure}
\includegraphics[width=\linewidth,clip=true]{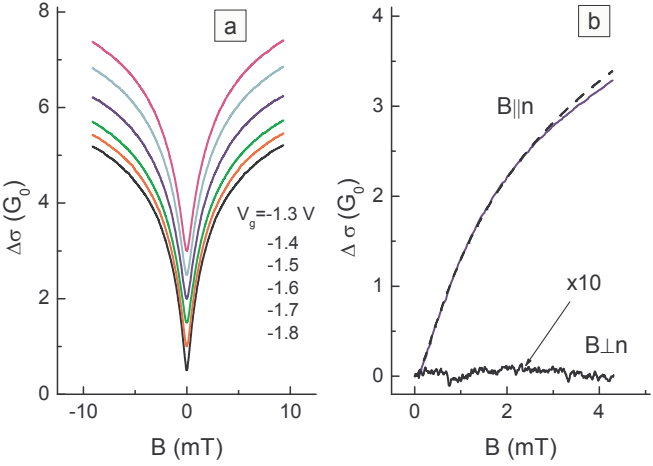}
\caption{(Color online) (a) The interference induced positive magnetoconductivity
for different gate voltages near the balance. For clarity, the curves are shifted in the vertical direction.
(b) The  magnetoconductivity  measured at $V_g=-1.5$~V for two orientations of the magnetic field.
Solid lines are the data, the dashed line is the the best fit by Eq.~(\ref{eq50}) with the
parameters $\alpha=1.9$, $\tau_\phi/\tau=85$. $T=1.35$~K for all the cases, $B_{tr}=8.4$~mT.
}\label{f5}
\end{figure}

Now we are ready to discuss quantitatively the low field negative
magnetoresistance caused by suppression of the weak localization. The
magnetic field dependences of $\Delta\sigma(B)$ at temperature $1.35$ K
for different gate voltages are presented in Fig.~\ref{f5}(a). At first
sight all the curves are the same  in the shape and   no peculiarity in
the balance at $V_g=-1.5$ V is observed. This, indeed, should be when
the interwell transition rate is low, $1/\tau_{12}\ll 1/\tau_\phi$, and
$\Delta\sigma(B)$ is the sum of the contributions from each well.

\begin{figure}
\includegraphics[width=\linewidth,clip=true]{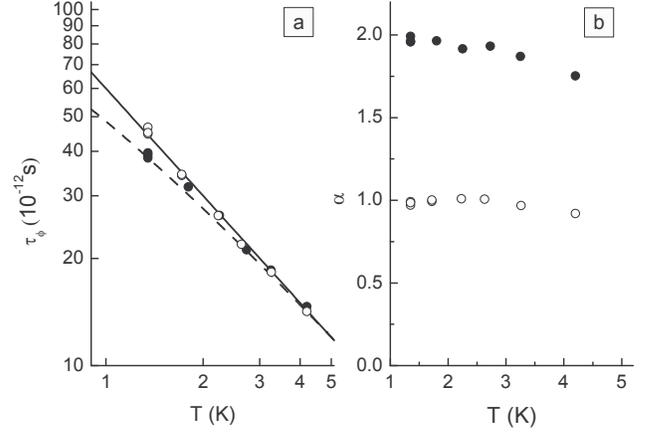}
\caption{(Color online) The temperature dependences of $\tau_\phi$ (a) and $\alpha$ (b) found from
the fit of the magnetoconductivity curves by Eq.~(\ref{eq50}). Solid symbols are obtained
at $V_g=-1.5~V$ when the structure is in the balance. Open symbols correspond to $V_g=-3.7~V$ when
the conductivity over the upper well is negligible. The solid and dashed lines are the functions
$6\times 10^{-11}/T$ and $(T/6.4\times10^{-11} +1/2\times10^{-10})^{-1}$, respectively.
}\label{f6}
\end{figure}

Closeness of the mobilities in the balance means closeness of transport
fields, $B_{tr1}= B_{tr2}=B_{tr}$ so that $\Delta\sigma(B)$ should be
described by Eq.~(\ref{eq10}), which in this case reduces to the
following
\begin{equation}
 \Delta\sigma(B)=\alpha\,G_0 {\cal H}\left(\frac{\tau_{\phi }}{\tau}\frac{B}{B_{tr}}\right),
\label{eq50}
\end{equation}
where $\alpha=2$. Therefore, let us start analysis from $V_g=-1.5$ V.
The result of the best fit within magnetic field range $(0-0.3)B_{tr}$
for $T=1.35$ K is shown in Fig.~\ref{f5}(b). It is evident that
Eq.~(\ref{eq50}) perfectly fits the data with prefactor which value is
really about two. The temperature dependences of the fitting parameters
$\tau_\phi$ and $\alpha$ are presented in Fig.~\ref{f6}. It is seen
that the $T$ dependence of $\tau_\phi$ is close to $1/T$ (some
deviation will be discussed below). Slight decrease of the prefactor
with the growing temperature  is sequence of the lack of the diffusion
regime due to the decrease of the $\tau_\phi$ to $\tau$
ratio\cite{Min00-1} from $85$ at $T=1.35$ K to $30$ at $T=4.2$ K.

\begin{figure}
\includegraphics[width=\linewidth,clip=true]{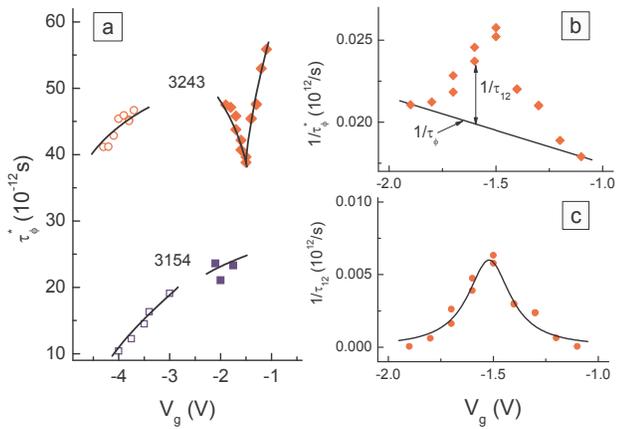}
\caption{(Color online) (a) The values of the fitting parameter $\tau_\phi^*$
plotted against the gate voltage for structures 3243 and 3154.
Solid symbols are obtained near the balance evident in the structures 3243 and
3154 at $V_g^b\simeq -1.5$~V and $-2.0$~V, respectively.
The open symbols correspond to the situation when only the lower quantum well contributes to
the conductivity. The lines are provided as a guide to the eye. (b) The $1/\tau_\phi^*$
values as a function of the gate voltage for structure 3243 near the
balance. (c) The gate voltage dependence of the transition rate for
structure 3243. Symbols are the experimental data, the line is
calculated from Eq.~(\ref{eq70}) with $m=0.067m_0$, $dn/dV_g=2.8\times
10^{11}$~cm$^{-2}/V$, $\tau_{12}^b=1.7\times 10^{-10}$~s, and
$\tau=5.5\times 10^{-13}$~s. }\label{f7}
\end{figure}

Let us now analyze the data in the vicinity of $V_g=-1.5$~V. Strictly
speaking, the use of Eq.~(\ref{eq50}) is not fully correct because
$B_{tr}$ and  $\tau_{\phi i}/\tau_i$  are different in the wells.
Nevertheless, before to discuss the workability of Eq.~(\ref{eq50}) for
obtaining of $\tau_\phi$ experimentally, let us consider the results of
such treatment. The gate voltage dependence of $\tau_\phi$ presented in
Fig.~\ref{f7}(a) by the diamonds exhibits the sharp minimum at
$V_g=-1.5$~V. To understand whether or not this minimum results from
the approximation used, we have simulated experimental situation. We
have calculated the magnetoresistance using Eq.~(\ref{eq10}) with the
values of $n_i$ and $\mu_i$ given in Fig.~\ref{f3}. The phase
relaxation times have been calculated according to
Ref.~\onlinecite{Nar02}. Then, these curves have been fitted by
Eq.~(\ref{eq50}) in the same manner as it has been done with the data.
It turns out that Eq.~(\ref{eq50}) well fits the calculated curves.
Therewith, the fitting parameter $\tau_\phi$ monotonically changes with
the changing gate voltage and  its value is close to $(\tau_{\phi
1}+\tau_{\phi 2})/2$ with an accuracy of $5$\% in the range of the
parameters corresponding to $V_g=-(1.3\ldots 1.7)$~V. This means that
the minimum in Fig.~\ref{f7} is not sequence of the fitting procedure.

It stands to reason that the minimum results from the interwell
transitions neglected in Eq.~(\ref{eq50}). Theoretically, their role in
the positive magnetoconductivity was considered in
Refs.~\onlinecite{Averkiev98} and \onlinecite{Raichev00}. However
attempts to use the expressions from these papers for the data fit have
failed. The values of $\tau_\phi$ and $\tau_{12}$ are obtained with
very large uncertainty.  This is because the shape of the
magnetoconductivity curve in actual case is mainly controlled by some
combination of $\tau_\phi$ and $\tau_{12}$ but not by $\tau_\phi$ and
$\tau_{12}$ individually. It becomes clear when considering the
parabolic (at $B\to 0$) and logarithmic (at $B\gg
B_{tr}\tau/\tau_\phi$) asymptotics of Eq.~(24) in
Ref.~\onlinecite{Raichev00} at $\tau_{12}> 2\,\tau_\phi$. They coincide
with corresponding asymptotics, which one can extract from
Eq.~(\ref{eq50}) by using $\alpha=2$ and
$\tau_\phi^\star=(1/\tau_\phi+1/\tau_{12})^{-1}$ instead of
$\tau_\phi$.\footnote{The numerical simulation shows that this
conclusion is valid over the whole magnetic field range including the
intermediate fields.} So, the sum $1/\tau_\phi+1/\tau_{12}$, but not
the rates $1/\tau_\phi$ and $1/\tau_{12}$ separately are experimentally
obtained when the interwell transitions are relatively rare.

According to Ref.~\onlinecite{Raichev00}, the interwell transition rate
resonantly depends on the difference between the Fermi energies in the
wells:
\begin{equation}
 \frac{1}{\tau_{12}}= \frac{1}{\tau_{12}^b} \frac{1}{1+[(E_F^1-E_F^2) \tau/\hbar]^2},
\label{eq60}
\end{equation}
where $1/\tau_{12}^b$ is the rate in the balance and
$E_F^i=\pi\hbar^2n_i/m$ is the Fermi energy in {\it i}-th well. Taking
into account the fact that $n_1$ is practically independent of $V_g$,
one obtains the following expression
\begin{equation}
 \frac{1}{\tau_{12}}=\frac{1}{\tau_{12}^b} \left[1+\left(\frac{\pi\hbar}
 {m}\frac{dn}{dV_g}(V_g-V_g^b)\tau\right)^2\right]^{-1},
\label{eq70}
\end{equation}
which can be directly applied to describe the data. Experimentally, the
rate $1/\tau_{12}$ near the balance can be found as the difference
between $1/\tau_\phi^\star$ and  $1/\tau_\phi$, if one supposes that
the interwell transitions do not contribute to $\tau_\phi^*$ at
$V_g=-1.8$~V and $V_g=-1.2$~V, and $1/\tau_\phi$ linearly depends on
$V_g$ between these gate voltages as shown in Fig.~\ref{f7}(b). The
results are presented in Fig.~\ref{f7}(c) by the circles. As seen the
experimental $V_g$ dependence of $1/\tau_{12}$ is really close to the
theoretical one calculated from Eq.~(\ref{eq70}) with $\tau=5.5\times
10^{-13}$~s found from the mobility value at $V=V_g^b$ and
$\tau_{12}^b= 1.7\times 10^{-10}$~s.

\begin{figure}
\includegraphics[width=\linewidth,clip=true]{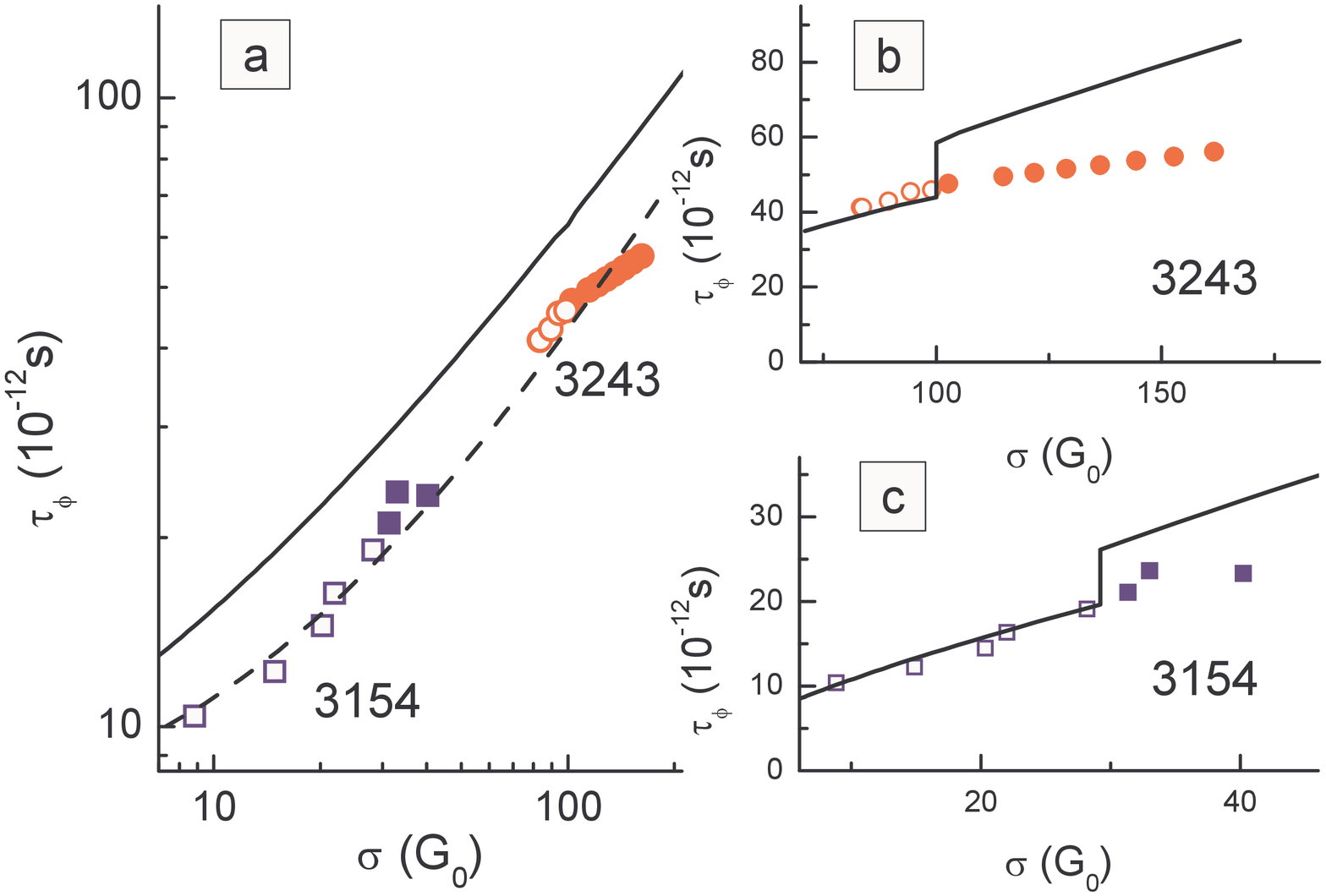}
\caption{(Color online) The conductivity dependence of the phase relaxation time
for the structure 3243 (circles) and
3154 (squares) at $T=1.35$~K. Solid symbols are obtained near the balance. Open
symbols correspond to the situation when only the lower quantum well contributes to
the conductivity. The solid line in the panel (a) is calculated according to Ref.~\onlinecite{Nar02}
for single quantum well, while the dashed line is provided as a guide to the eye. The lines in
panels (b) and (c) are expected dependence taking into account the
presence of the second well according to Eq.~(\ref{eq80}). }\label{f8}
\end{figure}

Thus, we obtain from the above analysis that the dephasing time
$\tau_\phi$ in the balance is about $5\times 10^{-11}$~s at $T=1.35$~K,
whereas the interwell transition time is approximately three times
larger, $\tau_{12}\simeq 1.7\times 10^{-10}$~s. The fact that
$\tau_\phi^*$ is the combination of $\tau_\phi$ inversely dependent on
the temperature and $\tau_{12}$, which is independent of the
temperature, explains the tendency to saturation of the $\tau_\phi^*$
vs $T$ dependence at low temperature [see the solid symbols and the
dashed curve in Fig.~\ref{f6}(b)].

It is thought that the role of the interwell transitions can be
independently estimated from the in-plane
magnetoconductivity.\cite{Raichev00,Min00-4} However, the detection of
this effect  is beyond accuracy of our experiment as readily
illustrated by Fig.~\ref{f5}(b). Its magnitude can be estimated as at
list $100$ times less than that at $B\parallel n$. This agrees with the
estimations made according to Ref.~\onlinecite{Raichev00}. They show
that the value of the in-plane magnetoconductivity with the parameters
of the structure investigated should be about $1000$ times less than
the value of the magnetoconductivity at $B\parallel n$.

It is impossible to find the values of $\tau_{\phi 1}$ and $\tau_{\phi
2}$ within the gate voltages range from $-3.5$~V to $-2.0$~V because
six independent parameters govern  the magnetoconductivity [see
discussion below Eq.(\ref{eq10})]. However, it can be easily done at
lower gate voltage, $V_g<-3.7$~V, when the conductivity and, thus, the
magnetoconductivity are determined by the lower well only. The
$\Delta\sigma$ vs $B$ curve for this case is well fitted by
Eq.~(\ref{eq50}) with the prefactor $\alpha$, which value is close to
$1$ [see Fig.~\ref{f6}(b)]. Therewith the temperature dependence of
$\tau_\phi$, as seen from Fig.~\ref{f6}(a), is close to $1/T$. The
values of $\tau_\phi$ within this gate voltage range are shown in
Fig.~\ref{f7}(a) by the open circles.

The same measurements were carried out on the structure 3154. The
balance in this structure occurs at $V_g=-2$ V, at the lower electron
density, $n\simeq 4.5\times 10^{11}$~cm$^{-2}$, and lower mobility,
$\mu\simeq 6500$~cm$^2$/Vs. The results differ by two issues. The value
of $\tau_\phi$ is less than that for the structure 3243. It seems
natural because  the conventional theory\cite{AA85,Nar02} predicts that
$\tau_\phi\propto \sigma/\ln{(\sigma/G_0)}$. The second point is that
the noticeable minimum at the balance is not observed [see
Fig.~\ref{f7}(a)]. It is explained by the relatively low interwell
transition rate, which is estimated\cite{Raichev00} as being about two
times lower than that in the structure 3243 and more than six times
lower than $1/\tau_\phi$ at lowest temperature.

As mentioned above, the dephasing time $\tau_\phi$ should mainly depend
on the conductivity, at least for the single well structure.  The open
symbols in Fig.~\ref{f8}(a) are the data obtained for both structures
for the case when the conductivity is determined by the one well alone.
The theoretical dependence\cite{Nar02} shown in the same figure
reasonably describes all these data. The quantitative agreement is not
to be anticipated because the expression used for the data treatment
does not take into account the magnetic field dependence of
$\tau_{\phi}$, the contributions of the second order corrections, and
the finiteness of the $\tau_\phi$ to $\tau$ ratio.

Let us again direct the reader's attention to the data obtained near
the balance. These data are presented in Fig.~\ref{f8} by the solid
symbols. Note, they are plotted against the conductivity per one well.
As seen from Fig.~\ref{f8}(a) the experimental $\tau_\phi$ values
obtained under the conditions, when one and both quantum wells are
occupied, fall on the common curve. This means that the inelastic
interaction of an electron in the one well with electrons in the other
one does not contribute noticeably to the dephasing rate. At first
glance the last seems very strange because the distance between the
wells is about screening length so that the interaction between the
carriers located in the different wells should not be significantly
less than that for carriers in the one well. On the other hand, the
inelastic interaction of an electron with electrons in the other well
is not sole effect influencing the dephasing rate. Another effect is an
additional screening of the interaction between the electrons in the
one well by the carriers located in the other one. The former effect
should lead to the $\tau_\phi$ decrease, while the second one to its
increase. Theoretical calculations, which take into account both these
effects, give the following expression for the dephasing time in a
double layer system:\cite{IgorAndIgor}
\begin{equation}
\frac{\tau_\phi^{(1)}}{\tau_\phi} = \left[1-\frac{1}{2} \frac{3\varkappa d+2}{(\varkappa d
+2)(\varkappa d +1)}\right]\frac{\ln{T\tau_\phi}}{\ln{T\tau_\phi^{(1)}}},
\label{eq80}
\end{equation}
where $\tau_\phi^{(1)}$ denotes the dephasing time of a single layer,
$d$ is the distance between the layers, and $\varkappa=1/r_0$. Using
the distance between the centers of the wells, $18$~nm, as the $d$
value and $r_0=5$~nm, we obtain approximately $25$ percent increase of
$\tau_\phi$ at $T=1.35$~K, which should occur according to
Eq.~(\ref{eq80}) when the upper quantum well starts to be occupied [see
lines in Figs.~\ref{f8}(b) and \ref{f8}(c)]. It may appear that this
change is not sufficiently large to be fixed experimentally. However,
as evident from Fig.~\ref{f7}(a) we reliably recognize the $20$~percent
resonant dip in the $\tau_\phi^*$~vs~$V_g$ dependence resulting from
the interwell transitions, but observe no change in $\tau_\phi$ of the
lower well when occupying the upper one out of resonance (see symbols
in Fig.~\ref{f8}). It should be noted that Eq.~(\ref{eq80}) takes into
account the interaction appearing in the singlet channel only. So, the
possible reason for disagreement between the experimental data and
calculation results is the neglecting of the contribution of the
multiplet (with respect to spin and well isospin) channel in the
dephasing rate. To  our knowledge the role of the {\it e-e} interaction
in the multiplet channel in the dephasing in the double quantum well
structures has not been studied yet in the literature.

Before concluding the paper, let us  note that our results are in
contradiction with those obtained in Ref.~\onlinecite{Pagnossin08}. In
the structures with the relatively high interwell transition rate, the
authors observe the anomalous $\sigma$ dependence of the dephasing
time, $\tau_\phi\sim \sigma^{-1}$. This observation is inconsistent not
only with the data in Fig.~\ref{f8} but  with the Fermi-liquid
model,\cite{Nar02} which predicts $\tau_\phi\sim \sigma$. The possible
reason is that the authors used the expression obtained in
Ref.~\onlinecite{Raichev00} for the case of identical wells, whereas
the noticeable classical positive magnetoresistance observed at all the
gate voltages  fairly indicates that the mobilities in the wells are
different.

\section{Conclusion}
We have studied the interference quantum correction in the gated double
quantum well Al$_x$Ga$_{1-x}$As/GaAs/Al$_x$Ga$_{1-x}$As
heterostructures. Analyzing the positive magnetoconductivity we have
obtained the interwell transition rate and the phase relaxation rate
under the conditions when one and two quantum wells are occupied. It
has been found that the interwell transition rate resonantly depends on
the difference between the electron densities in the wells in
accordance with the theoretical estimate. The central point of the
paper, however, is that the dephasing rate in the lower quantum well is
independent of whether the upper quantum well contributes to the
conductivity or not. This observation is inconsistent with the results
of simplest theory, which takes into account the inelasticity of the
{\it e-e} interaction in the singlet channel only, and predicts the
increase of the dephasing time in double layer structures as compared
with the single layer case. The further experimental and theoretical
investigations are needed to find the answer to the question of whether
the interaction in the multiplet channel or some other mechanism is
responsible for such the feature of the dephasing processes in double
quantum well structures.

\section*{Acknowledgments} We would like to thank I.~S.~Burmistrov and
I.~V.~Gornyi for illuminating discussions. This work has been supported
in part by the RFBR (Grant Nos 08-02-00662, 09-02-00789, and
10-02-00481).

%\bibliography{QuantumCorrections}

%merlin.mbs 2010-03-15 4.21a (PWD, AO, DPC)

\end{document}